# Femtosecond phase-transition in hard x-ray excited bismuth

*M. Makita [a], I. Vartiainen [a], I. Mohacsi [a], C. Caleman [b,c], A. Diaz [a], H. O. Jönsson [c], P. Juranić [a], N. Medvedev [d,e], A. Meents [b], A. Mozzanica [a], N. Opara [a], C. Padeste [a], V. Panneels [a], V. Saxena [b,f], M. Sikorski [g], S. Song [g], L. Vera [a], P. R. Willmott [a], P. Beaud [a], C.J. Milne [a], B. Ziaja-Motyka [b,h], and C. David [a]

(a) Paul Scherrer Institut, CH-5232 Villigen PSI, Switzerland
(b) CFEL, Deutsches Elektronen-Synchrotron DESY, 22607 Hamburg, Germany
(c) Department of Physics and Astronomy, Uppsala University, SE-751 24 Uppsala, Sweden
(d) Institute of Physics, Czech Academy of Sciences, 182 21 Prague 8, Czech Republic
(e) Institute of Plasma Physics, Czech Academy of Sciences, 182 00 Prague 8, Czech Republic
(f) Institute for Plasma Research, Bhat, Gandhinagar 382428, India
(g) Linac Coherent Light Source, SLAC National Accelerator Laboratory, 2575 Sand Hill Road, Menlo Park, California, 94025, USA
(h) Institute of Nuclear Physics, Polish Academy of Sciences, 31-342 Krakow, Poland

**[Abstract]**

**The evolution of the bismuth crystal structure upon excitation of its $A_{1g}$ phonon has been intensely studied with short pulse optical lasers. Here we present the first-time observation of a hard x-ray induced ultrafast phase transition in a bismuth single crystal, at high intensities (~$10^{14}$ W/cm$^2$). The lattice evolution was followed using a recently demonstrated x-ray single-shot probing setup. The time evolution of the (111) Bragg peak intensity showed strong dependence on the excitation fluence. After exposure to a sufficiently intense x-ray pulse, the peak intensity dropped to zero within 300fs, i.e. faster than one oscillation period of the $A_{1g}$ mode at room temperature. Our analysis indicates a nonthermal origin of a lattice disordering process, and excludes interpretations based on electron-ion equilibration process, or on thermodynamic heating process leading to a plasma formation.**

The typical response time of the internal microscopic degrees of freedom in a solid, such as the arrangement of the electrons and atoms ranges between few fs to few ps. Ultrashort laser pulses can excite materials on time scales that are faster than those response times, often revealing unique behaviours [1–5] and furthering the understanding of the interactions between electrons and lattice [6–8]. In the case of ultrafast melting, photoexcitation drives the material in a highly non-equilibrium state where the electrons are at very high temperature while the lattice is still cold. As the electrons thermalize, the lattice disorders in hundreds of femtoseconds due to a significant shift in the atomic potential energy surface (PES) driven by the excited electrons. Bismuth (Bi), well known for its Peierls distorted lattice structure, shows one important example of such a phase transition.

The crystal lattice of Bi can be readily excited by an ultrashort laser pulses to promote electronic excitations that typically trigger a coherent oscillation of the optical Γ-point $A_{1g}$ phonon mode [1–4,9]. The characteristic parameters of this oscillation, for example its frequency (~2.9THz [10,11]), are strongly dependent on the excitation intensity and are directly correlated with the out-of-equilibrium potential energy surface (PES) [12]. Results from a pioneering infrared-pump experiment [13] and subsequently from theoretical models [8], suggest the presence of a non-thermal melting process in Bi at an absorbed dose beyond its melting threshold. To date, however, neither experiments nor calculations have been able to describe the dynamics in a range beyond an absorbed dose of 1.2 eV/atom [9], which is still well below a fast ionisation regime. Furthermore, high time resolution structural studies sensitive to the excitation of phonon modes in the material is missing. It is therefore a fundamental interest to investigate the dynamical lattice response of Bi to intense electronic excitations, with high time resolutions.

To address this question, we study the ultrafast lattice dynamics of a bismuth bulk crystal, excited with a femtosecond hard x-ray pulse. The advantages of using hard x-rays, instead of optical lasers, are: (i) it allows to excite the material at high doses with negligible non-linear effects from electric and magnetic fields of the focused laser [14], thus a purely electronic response of the material could be obtained, and (ii) it creates a low electron density gradient within the pump penetration depth, thus a relatively homogeneous excitation profile within the probed depth. The technical difficulties typically associated with an x-ray pump and x-ray probe experiment using SASE pulses [15] were overcome by using a novel x-ray splitting setup [16]. As a result, an extremely high temporal resolution was achieved, limited only by the XFEL pulse duration. Importantly, this technique is free of any timing and spatial jitter between the pump and the probe, or between the probe beams.

The experiment was performed at the XCS station [17] of the Linac Coherent Light Source (LCLS) [18], using an x-ray photon energy of 5 keV, a nominal pulse energy of 2 mJ and a nominal pulse duration of 35 fs FWHM. The choice of the x-ray energy does not induce a resonant excitation of inner-shell electrons. The schematic arrangement of the transmission gratings is illustrated in Fig.

1a. The transmitted part of the x-ray pulse through the gratings acted as a 'pump', whilst those diffracted by the gratings 'probed' the sample at precise time delays determined by their extended optical paths relative to the transmitted pulse. On the sample, half of the probe beams are spatially overlapped with the pump pulse, while the other half was separated by 70 μm from the pumped region. More details about the setup can be found in the "methods' and in ref. [16]. To ensure the spatial overlap, spot sizes of the pump and the probe pulses were focused to a FWHM of 35 ± 5 μm and 12.5 ± 2.5 μm, respectively. The sample was oriented so as to direct the Bragg reflections of all the pulses onto a 2D detector.

Figure 2a & 2b show typical single-shot images of the Bi (111) Bragg peak reflections which is sensitive to the $A_{1g}$ phonon mode, as recorded on the detector. Diffraction patterns with and without the pump pulse are compared (Fig. 2a & 2b). For both images, the group of signals on the left-hand side, marked as "reference", probed the unpumped region of the sample; while that marked as "probe" on the right, probed the excited area of the sample. The delay time increases as the signal distance increase from the (blocked) pump pulse at the centre of the image. The flare caused by the pump beam was fitted with a 1D Gaussian profile for background removal (Fig. 2c).

Figure 3 shows the time evolution of the (111) reflection recorded for different x-ray pump fluences, with the relative pump incidence at time 0 (centre of the pump pulse). The fluence levels are 2.46 J/cm$^2$, 1.48 J/cm$^2$, and 0.91 J/cm$^2$, corresponding to the average absorbed dose values of 3.5 eV/atom, 2.1 eV/atom, and 1.3 eV/atom, respectively. On average, 50 ± 10 selected shots for each fluence level were used. We emphasize here that the signal dynamics seen in Fig. 3 cannot be due to x-ray intensity and spectral fluctuation, or due to crystal imperfections, since the time sequences were taken from the same pulse in single shot and from the same region of the crystal.

The dynamics at the lowest absorption dose of 1.3 eV/atom (at 0.91 J/cm$^2$) in our case could be considered analogous to the highest dose cases presented in ref. [13] (1.2 eV/atom for Bi) and in ref. [19] (0.35 eV/atom for Sb) – implying that the PES map at this x-ray dose should be highly anharmonic and described by a single-minimum-well. For such case it is expected that the disintegration of the lattice follows a relatively direct path, possibly by a direct anharmonic coupling of the initially excited $A_{1g}$ motion coordinate to that of the degenerate $E_g$ optical mode [8,13].

For the absorbed dose of 2.1 eV/atom and 3.5 eV/atom, the intensity drops completely within ~300 fs, faster than one oscillation period of the unperturbed Bi $A_{1g}$ phonon mode (342 fs, 2.92 THz) [10–12]. This is also highlighted in the inset graph in Fig.3, which compares our observation to previous Bi (111) diffraction experiments in which bond softening or lattice damage at later times were observed [2,4,9]. Our signal decay in less than half a picosecond eliminates the possibility of thermal melting process. In the thermal regime, dramatic changes in the crystal structure can only be observed after the electron-ion equilibration time which is of the order of ps [20,21]. We also note that the observed femtosecond transition triggered by $10^{13} \sim 10^{14}$ W/cm$^2$ pump intensity, excludes the possibility of collisional heating through photoexcited electrons [22] as a dominant heating process. This conclusion has been confirmed by the ionization degree estimates obtained with the XATOM code [23], and with the non-local thermodynamic-equilibrium (non-LTE) radiation transfer code, CRETIN [24]. At the highest considered x-ray dose (3.5 eV/atom) both calculations indicate an average ionization level of less than 1 per atom at 100 fs after the excitation event, confirming that a plasma formation regime was not reached.

An intermediate signal plateau present for all fluence cases shown in Fig. 3 (shaded green area) has been observed at a comparable timescale in x-ray irradiated diamond, within its transient transmission of optical pulses [25]. An exponential fitting to extrapolate the decay time constant is possible for reflections from lattice insensitive to optical phonon-modes [5,13]. In our case, however, due to the (111) reflection sensitivity to the $A_{1g}$ phonon mode, we cannot uniquely exclude the possible origin of the plateau in Fig. 3 from a damped oscillatory motion along the distortion coordinate [26,27]. In such case the major initial atomic potential shift would be along the (111) direction, leading to minima of diffracted intensity when Bi atomic potential pass through or near the lower symmetry equilibrium position of the parent cubic lattice. Evidence of such process could be obtained by performing dedicated experiments with a longer time delay window.

A possible interpretation of the fast (111) signal drop includes non-thermal melting during which a modification of interatomic potential due to high electronic excitation induces a lattice instability [13,28]. Fig. 3 also shows that both the amplitude and the slope of the initial signal drop exhibit a fluence dependence – suggesting that a displacive excitation of coherent phonons (DECP) [12] could have been triggered by x-rays. Other scenarios such as a decrease of the diffracted intensity because of thermal excitation of the lattice (Debye-Waller) would exhibit a fluence dependence only in the amplitude. The theoretical modelling of the DECP process is primarily dependent on the pump pulse duration (shorter than the phonon oscillation period) and on the absorbed power density. The absorbed energy could be assigned either to the excited electron density $n(t)$, or to the excited electron temperature $\Delta T_e$, a more likely scenario for hard x-ray excitation where energetic core- electrons are often produced. Employing either parameter as a dominant source results in a similar conclusion, that is the sensitivity of the material to the PES displacement is strongly dependent on either $n(t)$, or $\Delta T_e(t)$. Our excitation parameters and the observation satisfy the condition for DECP, and we speculate that this coherent lattice motion could be the precursor to the observed phase transition on such a rapid timescale. Its implication would be a strong anharmonic coupling between the $A_{1g}$ and the $E_g$ optical modes leading to a structure disordering on the timescale less than one period of the $A_{1g}$ mode, as discussed also in ref. [8]. The physical processes of such transition dynamics initiated by hard x-rays, are therefore most likely relevant for materials with $A_1$ optical modes in general.

In conclusion, we have performed the first-time observation of x-ray induced lattice disordering in Bi occurring within less than 300 fs. The (111) Bragg peak dynamics indicates that the phase transition proceeds as a complex multistep process. Our data set an important benchmark for future experiments and modelling of the hard x-ray induced ultrafast phase-transition in Bi, which should reproduce the observed fluence dependence. For that purpose, there is a strong need for a development of dedicated theory models capable to follow such complex non-equilibrium dynamics in time. The promising candidates are currently developed time-dependent Density Functional Theory or Hartree-Fock based schemes.

**[Acknowledgement]**

We thank Dr. Steve Johnson, Dr. Stephen Fahy, and Dr. Eamonn Murray for insightful advices and discussions. The research leading to these results has received funding from the European Community's Seventh Framework Programme (FP7/2007-2013) under grant agreement no. 290605 (PSI-FELLOW/COFUND). Partial financial support from the Czech Ministry of Education (Grants

LG15013 and LM2015083) is acknowledged by N. Medvedev. C. Caleman thanks Swedish Research Council (2013-3940) and Helmholtz Association, through Center for Free-Electron Laser Science at DESY. Preliminary experiments were conducted at the cSAXS beamline at the Swiss Light Source, Paul Scherrer Institut, Villigen, Switzerland, as preparation for the experiment at the FEL. Use of the Linac Coherent Light Source (LCLS), SLAC National Accelerator Laboratory, is supported by the U.S. Department of Energy, Office of Science, Office of Basic Energy Sciences under Contract No. DE-AC02-76SF00515.

**[Figures]**

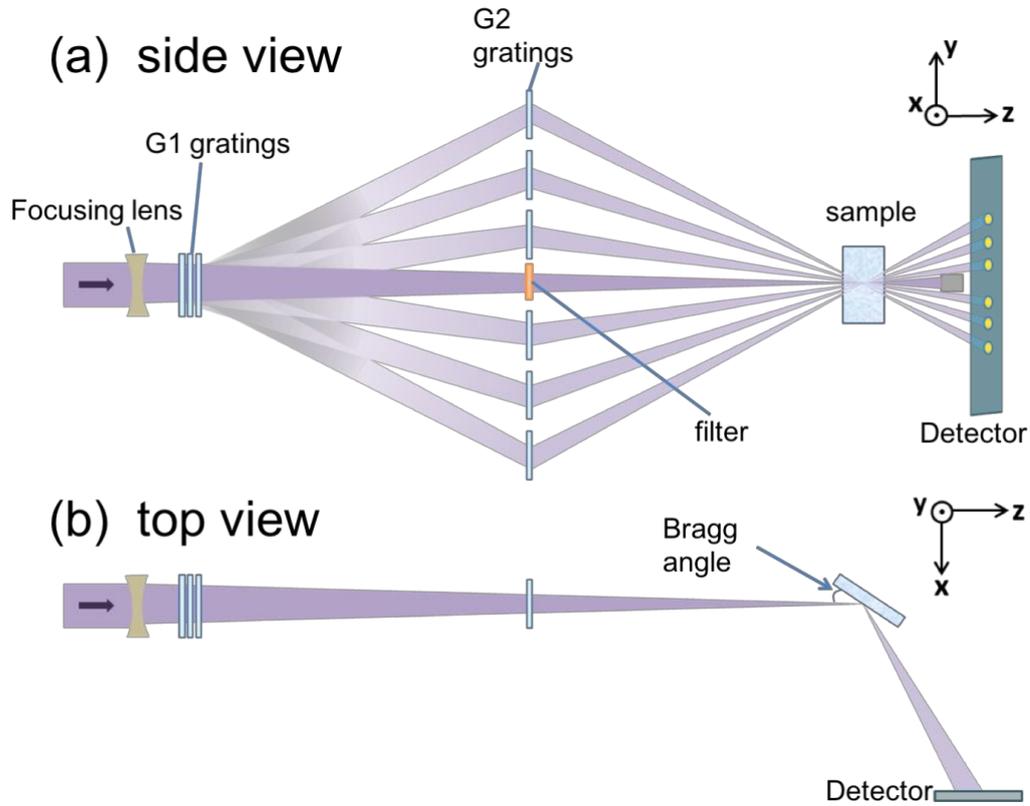

Figure 1. (a) Y-Z plane view of the setup. The label 'G1 gratings' denotes a stack of 10 different diamond gratings. Only three are shown here for simplicity. Each of these gratings diffracts a small portion of the incoming x-ray pulse in the y-z plane, at varying angles defined by their pitch. The diffracted pulses are then re-diffracted back by the G2-gratings, to overlap with the transmitted primary pulse within ±3 µm precision. (b) X-Z plane view of the setup. All the transmitted and diffracted pulses satisfy the Bragg condition of the sample in the x-z plane. The pulses that diffract from the sample are separated along the y-axis and aligned along the x-axis. They are then recorded on a 2D detector (JUNGFRAU).

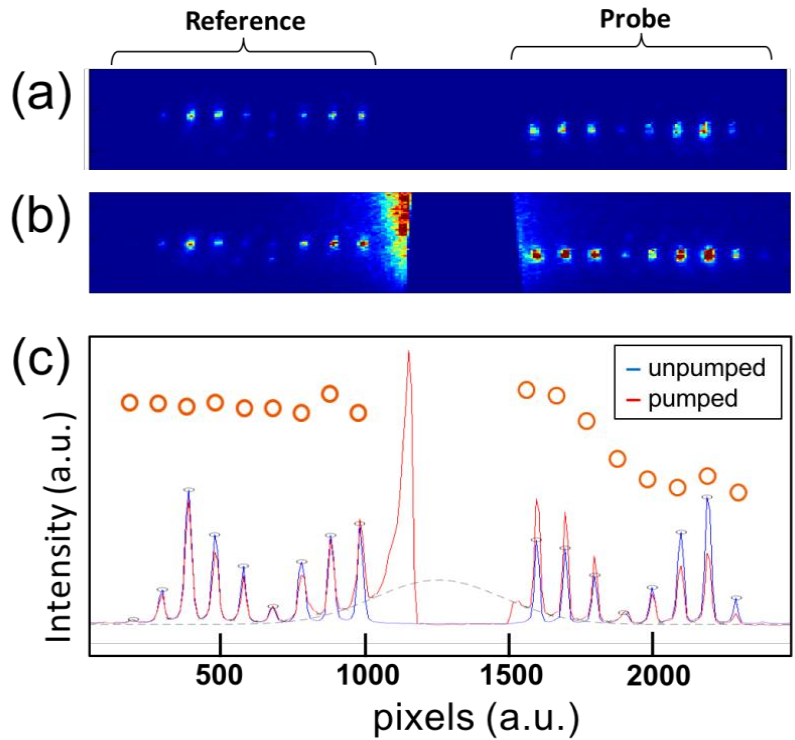

Figure 2. (a) & (b) Typical raw images showing (a) unpumped, and (b) pumped events of Bragg reflection signals. ((b) The pump pulse is blocked in front of the detector). The integrated plots of these images are shown in (c), red (pumped) and blue (unpumped) lines. The time delay increases symmetrically left and right from the pump-pulse at the centre. A dashed grey line indicates the 1D Gaussian background level in the pumped case. After background removal, the pumped signals are then integrated and normalised by their counterpart from unpumped signals. The normalised signals are plotted as red open circles.

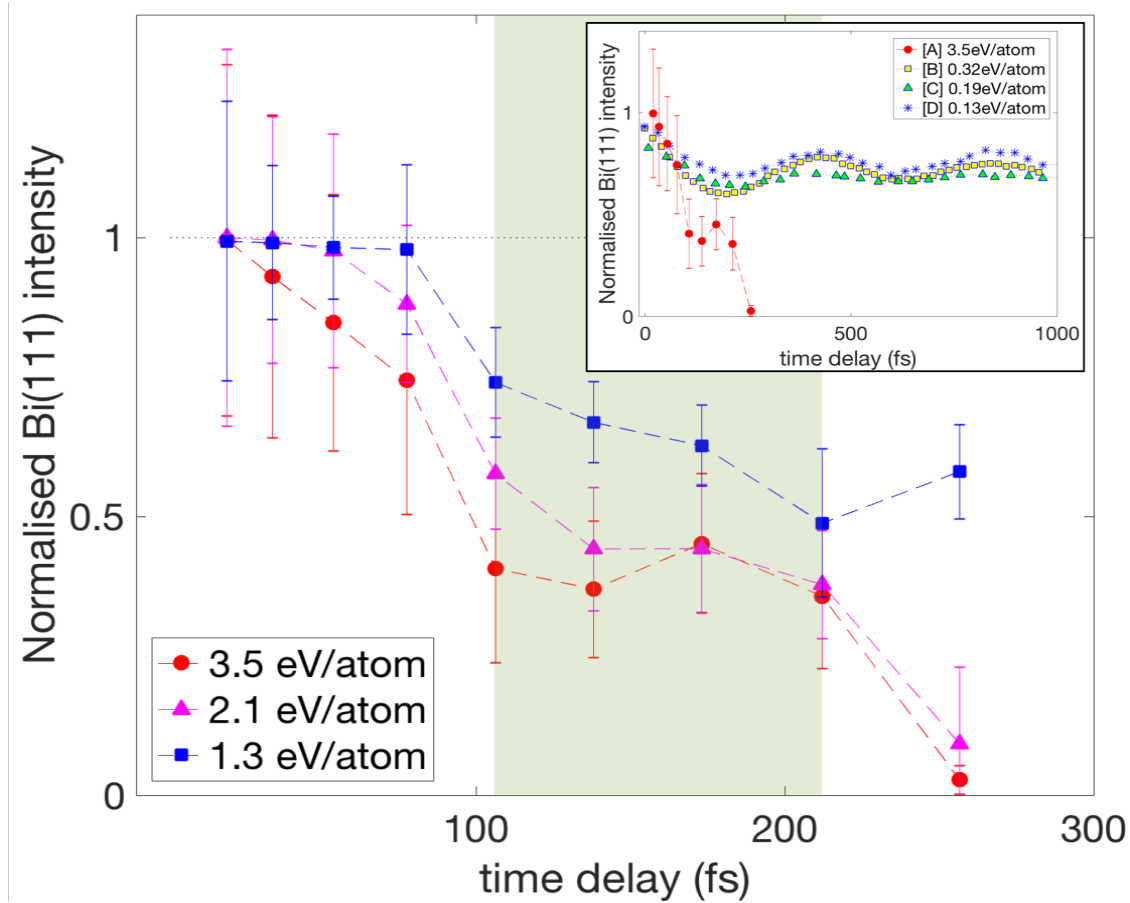

Figure 3. Temporal evolution of the normalised signals from x-ray irradiated Bi crystal at three different absorbed doses: 3.5 eV/atom (red circles), 2.1 eV/atom (magenta triangles) and 1.3 eV/atom (blue squares). Error bars are taken from standard deviation. Dashed lines are guides for the eye. The green shaded area marks roughly the plateau region discussed in the text. The inset graph shows the comparison of the case with [A] 3.5 eV/atom absorbed dose from 5 keV x-ray pulse, with the extrapolated data from previous reports on optical excitation of Bi at various absorbed doses: [B] Harmand et al.[6], [C] Fritz et al.[2], and [D] Johnson et al.[3].

**[Methods]**

The schematic arrangement of the transmission gratings is illustrated in Fig. 1a. A set of 10 linear gratings (G1 gratings), made of diamond, was placed 0.3 m downstream from the beamline exit window. A set of 20 linear gratings (G2 gratings), made of Ir and $SiO_2$, was positioned 3.3 m downstream of the G1 gratings to receive the diffracted beams from the G1 gratings. All the gratings were fabricated by means of electron-lithography and reactive ion etching, at the Laboratory for Micro- and Nanotechnology, Paul Scherrer Institut. The sample was positioned 6.6 m downstream of the G1 gratings, accepting both the transmitted (pump) beam and 20 diffracted (probe) beams from the G2 gratings. On the sample, half of the probe beams spatially overlap with the pump pulse, while the other half is separated by 70 μm to the unpumped region. In this way, the whole sequence of pumped and unpumped signals are obtained simultaneously, from a single x-ray pulse. In this experiment, the grating efficiencies are enhanced by factors of 2 ~ 5 compared to those reported earlier. The grating pitches were optimized to sequentially increase the delay timings with precisely defined intervals of 20 fs to 50 fs. This small time steps were continued up to ~ 300fs delay, with the probe focus on early-onset of the damage to the crystal.

The Bragg angle for the (111) reflection in Bi at 5 keV is 18 ° with a measured rocking curve width of 0.2°. The varying incident angles in y-z plane of the pump and probe beams do not affect the Bragg condition, because the diffraction angle from the sample is only sensitive to the angle in the x-z plane, which is perpendicular to the gratings' diffraction plane (Fig. 1b). Each pulse are then incident at the detector at spatially separated locations due to the angular separations in the y-z planes created by the diffraction gratings, thus retaining the delay-time information. Fresh sample surface was used for every shot.

The Bragg reflected probe beam signals were recorded using the JUNGFRAU detector, an integrating two-dimensional pixel detector comprised of arrays of 75 × 75 $μm^2$ pixels, with single photon sensitivity over a dynamic range of $2.5 \times 10^4$ for 5 keV photons. More details about the JUNGFRAU detector can be found in[1].

The intensity of each diffracted peak on the detector was determined by integrating the counts up to half the distance to the neighbouring peaks. The integrated peak intensities were then normalised against the unpumped counterparts, which were then rescaled using the normalised and averaged "reference" signals.

The size of the pump beam spot was determined by observing the attenuated x-ray beam on a YAG screen at the sample position, using a microscope lens and a Charge Coupled

Device (CCD) detector. The different spot sizes of the pump and the probe pulses were achieved by using Be focusing lenses positioned at two places. The first set was placed upstream of the G1 gratings and therefore affected the focusing of the direct pump beam and the probe beams. The other set was placed on the optical axis at the G2 grating position (not shown in the figure) and only affected the focusing of the pump pulse. This arrangement allowed us to choose the size of the pump and the probe beams independently.

The unattenuated pump pulse energy at the sample position (after passing through the gratings and the focusing lenses) was measured with a calorimeter to be ~100 μJ, which is ~5% of the nominal total emitted pulse energy. In order to vary the fluence level, but not the probe intensities, the pump pulse was attenuated using 10 μm and 20 μm Al foils placed on the optical axis of the transmitted beam, close to the G2 grating (Fig 1a). With these conditions, the pump beam fluence varied from 0.9 to 2.5 J/cm$^2$. The energy of each probe beam was 0.01 to 0.2 % of the pump beam at the sample position. In the absence of the pump beam, the crystal structure remains intact, confirming that the probe pulses are sufficiently weak to not damage the crystal.

**[Scattering factor and ionisation degree estimate with XATOM]**

Expected average photoionization degree of atoms within the irradiated Bi bulk was calculated with XATOM code[2] which is an integrated toolkit for x-ray atomic physics. Atomic data for all individual electronic configurations, including multiple-hole states of arbitrary atomic species, as well as the cross sections and rates of x-ray-induced atomic processes, can be calculated within the Hartree-Fock-Slater model. Figure below shows the calculated scattering factors of bismuth at different ionisation degrees.

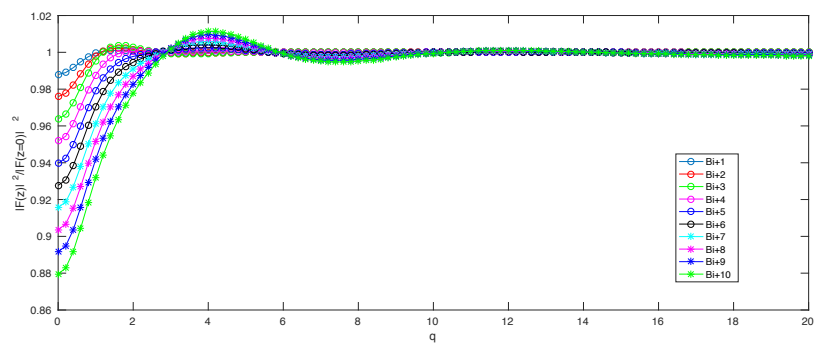

Figure (i): The scattering factors for various Bi ions. The plot suggests that the average ionization degree in Bi bulk would have to increase significantly within the 300 fs timescale, in order to explain the observed Bragg peak decay seen in Fig. 3 of the main manuscript as due to the scattering factor decrease.

**[Ionization degree calculation with CRETIN]**

CRETIN[3] is a 1D, 2D, and 3D non-local thermodynamic equilibrium (NLTE) atomic kinetics and radiation transport code which follows the time evolution of atomic populations and electron distributions as photons and electrons interacts with a plasma. Total average ionization degree, as well as the electron and ion temperatures of irradiated bismuth for an extended time period up to 100fs were calculated. The degree of ionisation shown below was estimated for the highest fluence case.

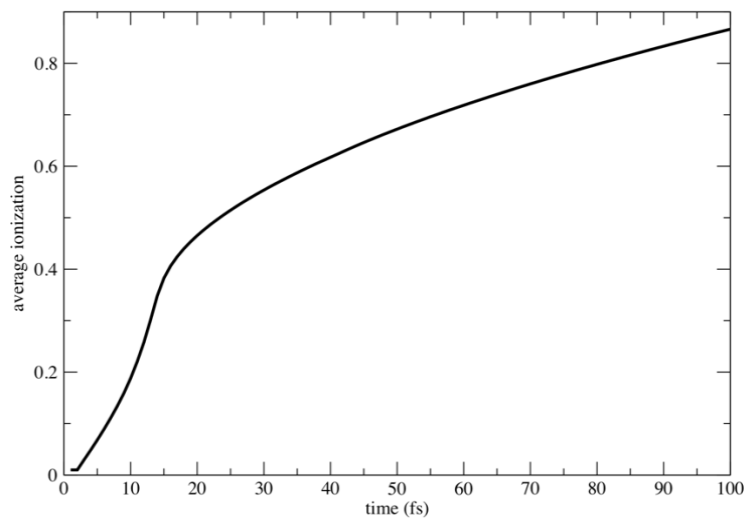

Figure (ii): Average ionization degree per atom of bismuth for the highest fluence case. Time zero denotes the maxima of the gaussian pump pulse. The average ionisation degree is less than 1, at 100 fs after the laser incidence, indicating that the sample will not reach the plasma state on the timescale of < 300 fs analysed in the manuscript.